\documentclass[a4paper,UKenglish]{lipics-cust}
 
\usepackage{complexity}
\usepackage{microtype}
\usepackage{soul}
\usepackage{placeins}

\usepackage{breakurl}

\usepackage{expl3, xparse, siunitx}
\ExplSyntaxOn
\NewDocumentCommand { \calcnum } { O{} m }
  { \num [  round-mode=figures , round-precision=3 , group-separator={.}, output-decimal-marker={,}, group-digits=integer] { \fp_to_decimal:n {#2} } }
\ExplSyntaxOff

\newcommand{\gtime}[1]{{\calcnum{#1 / 1000}}}
\newcommand{\ctime}[1]{\calcnum{#1 / 1000}}
\newcommand{\nexp}[1]{\calcnum{#1 / 1000000} $\times 10^6$}

\title{Computing Treewidth on the GPU}

\author[1]{Tom C. van der Zanden}
\author[1,2]{Hans L. Bodlaender}
\affil[1]{Department of Computer Science, Utrecht University, Utrecht,
The Netherlands\\
  \texttt{T.C.vanderZanden@uu.nl}}
\affil[2]{Department of Mathematics and Computer Science, Eindhoven University of
Technology, Eindhoven, The Netherlands
\\
  \texttt{H.L.Bodlaender@uu.nl}}
\authorrunning{T.\,C. van der Zanden and H.\,L. Bodlaender} 

\Copyright{Tom C. van der Zanden and Hans L. Bodlaender}


\begin{document}

\maketitle

\begin{abstract}
We present a parallel algorithm for computing the treewidth of a graph on a GPU. We implement this algorithm in OpenCL, and experimentally evaluate its performance. Our algorithm is based on an $O^*(2^{n})$-time algorithm that explores the elimination orderings of the graph using a Held-Karp like dynamic programming approach. We use Bloom filters to detect duplicate solutions.

GPU programming presents unique challenges and constraints, such as constraints on the use of memory and the need to limit branch divergence. We experiment with various optimizations to see if it is possible to work around these issues. We achieve a very large speed up (up to $77\times$) compared to running the same algorithm on the CPU.
\end{abstract}

\section{Introduction}
Treewidth is a well known graph parameter that measures how `tree-like' a graph is. The fact that many otherwise hard graph problems are linear time solvable
on graphs of bounded treewidth \cite{blsurvey} has been exploited in many theoretical and practical applications. For such applications, it is important to
have efficient algorithms, that given a graph, determine the treewidth and find tree decompositions with optimal (or near-optimal) width.

The interest in practical algorithms to compute treewidth and tree decompositions is also illustrated by the fact that both the PACE 2016 and
PACE 2017 challenges \cite{PACEcall} included treewidth as one of the two challenge topics. Remarkably, while most tracks in the PACE 2016 challenge
attracted several submissions \cite{PACEresults}, there were no submissions for the call for GPU-based programs for computing treewidth. 
Current sequential exact algorithms for treewidth are only practical when the treewidth is small (up to 4, see \cite{HeinK11}), or when the graph is
small (see \cite{gogate,ExactAlgos,bestTW,dowthesis,FPBB}). As computing treewidth is NP-hard, an exponential growth of the running time is to be expected;
unfortunately, the exact FPT algorithms that are known for treewidth are assumed to be impractical; e.g., the algorithm of \cite{Bodlaender96} has a running time
of $2^{O(k^3)} n$.
This creates the need for good parallel algorithms, as parallelism can help to significantly speed up the algorithms, and thus deal with larger graph sizes.

In this paper, we consider a practical parallel exact algorithm to compute the treewidth of a graph and a corresponding tree decomposition. The starting point of
our algorithm is a sequential algorithm by Bodlaender et al.~\cite{ExactAlgos}.
This algorithm exploits a characterization of treewidth in terms of the width of an {\em elimination ordering}, and gives a dynamic programming algorithm
with a structure that is similar to the textbook Held-Karp algorithm for TSP \cite{HeldK62}.

Prior work on parallel algorithms for treewidth is limited to one paper, by Yuan \cite{FPBB}, who implements a branch and bound algorithm for treewidth on
a CPU with a (relatively) small number of cores. With the advent of relatively inexpensive consumer GPUs that offer more than an order of magnitude more computational power than their CPU counterparts, it is very interesting to explore how exact and fixed-parameter algorithms can take advantage of the unique capabilities of GPUs. We take a first step in this direction, by exploring how treewidth can be computed on the GPU.

Our algorithm is based on the elimination ordering characterization of treewidth. Given a graph $G=(V,E)$, we may \emph{eliminate} a vertex $v\in V$ from $G$ by removing $v$ and turning its neighborhood into a clique, thus obtaining a new graph. One way to compute treewidth is to find an order in which to eliminate all the vertices of $G$, such that the maximum degree of each vertex (at the time it is eliminated) is minimized. This formulation is used by e.g. \cite{gogate} to obtain a (worst-case) $O^*(n!)$-time algorithm. However, it is easy to obtain an $O^*(2^n)$-time algorithm by applying Held-Karp style dynamic programming as first observed by Bodlaender et al. \cite{ExactAlgos}: given a set $S\subseteq V$, eliminating the vertices in $S$ from $G$ will always result in the same intermediate graph, regardless of the order in which the vertices are eliminated (and thus, the order in which we eliminate $S$ only affects the degrees encountered during its elimination). This optimization is used in the algorithms of for instance  \cite{bestfirst} and \cite{FPBB}.

We explore the elimination ordering space in a breadth-first manner. This enables efficient parallelization of the algorithm: during each iteration, a wavefront of states (consisting of the sets of vertices $S$ of size $k$ for which there is a feasible elimination order) is expanded to the wavefront of the next level, with each thread of the GPU taking a set $S$ and considering which candidate vertices of the graph can be added to $S$. Since multiple threads may end up generating the same state, we then use a bloom filter to detect and remove these duplicates.

To reduce the number of states explored, we experiment with using the minor-min-width heuristic \cite{gogate}, for which we also provide a GPU implementation. Whereas normally this heuristic would be computed by operating on a copy of the graph, we instead compute it using only the original graph and a smaller auxiliary data structure, which may be more suitable for the GPU. We also experiment with several techniques unique to GPU programming, such as using shared/local memory (which can best be likened to the cache of a CPU) and rewriting nested loops into a single loop to attempt to improve parallelism.

We provide an experimental evaluation of our techniques, on a platform equipped with a Intel Core i7-6700 CPU (3.40GHz) with 32GB of RAM (4x8GB DDR4), and an NVIDIA GeForce GTX 1060 with 6GB GDDR5 memory (Manufactured by Gigabyte, Part Number \texttt{GV-N1060WF2OC-6GD}). Our algorithm is implemented in OpenCL (and thus highly portable). We achieve a very large speedup compared to running the same algorithm on the CPU.

\section{Preliminaries}

\paragraph*{Treewidth.} For a detailed description of treewidth and its characterization, we refer to \cite{cygan}. Our algorithm is based on the $O(2^{n}nm)$-time algorithm of Bodlaender et al. \cite{ExactAlgos}. Though the characterization in terms of tree decomposition is more common, we recall only the characterization in terms of elimination orderings that is used by this algorithm:

Let $G=(V,E)$ be a graph with vertices $v_1,\ldots v_n$. An \emph{elimination ordering} is a permutation $\pi:V\to \{1,\ldots, n\}$ of the vertices of $G$. The \emph{treewidth} of $G$ is defined as $\min_\pi \max_v |Q(\{u\in V \mid \pi(u)<\pi(v)\}, v)| $, where $Q(S,v)$ is the set of vertices $\{u\in V\setminus S \mid \textrm{there is a path } v,p_1,\ldots,p_m,u \textrm{ such that } p_1,\ldots,p_m\in S \}$, i.e., $Q(S,v)$ is the subset of vertices of $V\setminus S$ reachable from $v$ by paths whose internal vertices are in $S$.

An alternative view of this definition is that given a graph $G$, we can \emph{eliminate} a vertex $v$ by removing it from the graph, and turning its neighborhood into a clique. The treewidth of a graph is at most $k$, if there exists an elimination order such that all vertices have degree at most $k$ at the time they are eliminated.

\paragraph*{GPU Terminology.} Parallelism on a GPU is achieved by executing many \emph{threads} in parallel. These threads are grouped into \emph{warps} of 32 threads. The 32 threads that make up a warp do not execute independently: they share the same program counter, and thus must always execute the same ``line'' of code (thus, if different threads need to execute different branches in the code, this execution is serialized - this phenomenon, called \emph{branch divergence}, should be avoided). The unit that executes a single thread is called a \emph{CUDA core}.

We used a GTX1060 GPU, which is based on the Pascal architecture \cite{whitepaper1080}. The GTX1060 has 1280 CUDA cores, which are distributed over 10 Streaming Multiprocessors (SMs). Each SM thus has 128 CUDA cores, which can execute up to 4 warps of 32 threads simultaneously. However, a larger number of warps may be assigned to an SM, enabling the SM to switch between executing different warps, for instance to hide memory latency.

Each SM has 256KiB\footnote{A \emph{kibibyte} is $2^{10}$ bytes.} of register memory (which is the fastest, but which registers are addressed must be known at compile time, and thus for example dynamically indexing an array stored in register memory is not possible), 96KiB of shared memory (which can be accessed by all threads executing within the thread block) and 48KiB of L1 cache.

Furthermore, we have approximately 6GB of global memory available which can be written to and read from by all threads, but is very slow (though this is partially alleviated by caching and latency hiding). Shared memory can, in the right circumstances, be read and written much faster, but is still significantly slower than register memory. Finally, there is also texture memory (which we do not use) and constant memory (which is a cached section of the global memory) that can be used to store constants that do not change over the kernel's execution (we use constant memory to store the adjacency lists of the graph).

Shared memory resides physically closer to the SM than global memory, and it would thus make sense to call it ``local'' memory (in contrast to the more remote global memory). Indeed, OpenCL uses this terminology. However, NVIDIA/CUDA confusingly use ``local memory'' to indicate a portion of the global memory dedicated to a single thread.

\section{The Algorithm}

\subsection{Computing Treewidth}

Our algorithm works with an iterative deepening approach: for increasing values of $k$, it repeatedly runs an algorithm that tests whether the graph has treewidth at most $k$. This means that our algorithm is in practice much more efficient than the worst-case $O^*(2^n)$ behavior shown by \cite{ExactAlgos}, since only a small portion of the $2^n$ possible subsets may be feasible for the target treewidth $k$. A similar approach (of solving the decision version of the problem for increasing values of $k$) was also used by Tamaki \cite{tamaki}, who refers to it as \emph{positive-instance driven dynamic programming}.

This algorithm lends itself very well to paralellization, since the subsets can be evaluated (mostly) independently in parallel. This comes at the cost of slightly reduced efficiency (in terms of the number of states expanded) compared to a branch and bound approach (e.g. \cite{dowthesis,FPBB,bestTW}) since the states with treewidth $< k-1$ are expanded more than once. However, even a branch and bound algorithm needs to expand all of the states with treewidth $k-1$ before it can conclude that treewidth $k$ is optimal, so the main advantage of branch and bound is that it can settle on a solution with treewidth $k$ without expanding all such solutions (of width $k$).

\begin{lstlisting}[caption={Algorithm for computing treewidth. Note that lines 7--19 compute the degree of $v$ in the graph that remains after eliminating the vertices in $S$.},label=list:8-6,captionpos=t,abovecaptionskip=-\medskipamount,mathescape=true,tabsize=2,numbers=left]
for k:=0 to n-1 do 
	inp:={$\emptyset$};
	for i:= 0 to n-k-2 do
		outp = {};
		foreach set $S$ in inp do		
			foreach vertex $v \not \in S$ do
				stack := {};
				degree := 0;
				push $v$ to stack;
				while stack $\not = \hspace{0.3em} \emptyset$ do
					pop vertex $u$ from stack;
					foreach unvisited neighbor $w$ of $u$ do
						mark $w$ as visited;
						if $w \in S$
							push $w$ to stack;
						else
							degree := degree+1;
					endforeach
				endwhile				
				if degree $\leq \hspace{0.3em} k$
					outp := outp $\cup \hspace{0.5em} \{S\cup \{v\}\}$;
			endforeach
		endforeach
		inp := outp
	endfor
	if inp $\not = \hspace{0.3em} \emptyset$
		report the treewidth of $G$ is $k$;
endfor
\end{lstlisting}\label{lst:basicalgo}

To test whether the graph has treewidth at most $k$, we consider subsets $S\subseteq V$ of increasing size, such that the vertices of $S$ can be eliminated in some order without eliminating a vertex of degree $> k$. For each $k$, the algorithm starts with an input list (that initially contains just the empty set) and then forms an output list by for each set $S$ in the input list, attempting to add every vertex $v\not \in S$ to $S$, which is feasible only if the degree of $v$ in the graph that remains after eliminating the vertices in $S$ is not too large. This is tested using a depth first search. Then, the input and output lists are swapped and the process is repeated. If after $n$ iterations the output list is not empty, we can conclude that the graph has treewidth at most $k$. Otherwise, we proceed to test for treewidth $k+1$. Pseudocode for this algorithm is given in Listing \ref{lst:basicalgo}.

We include three optimizations: first, if $C\subseteq V$ induces a clique, there is an elimination order that ends with the vertices in $C$ \cite{ExactAlgos}. We can thus precompute a maximum clique $C$, and on line 7 of Lisiting \ref{lst:basicalgo}, skip any vertices in $C$. Next, if $G$ has treewidth at most $k$ and there are at least $k+1$ vertex-disjoint paths between vertices $u$ and $v$, we may add the edge $uv$ to $G$ without increasing its treewidth \cite{disjointpaths}. Thus, we precompute for each pair of vertices $u,v$ the number of vertex-disjoint paths between them, and when testing whether the graph has treewidth at most $k$ we add edges between all vertices which have at least $k+1$ disjoint paths (note that this has diminishing returns, since in each iteration we can add fewer and fewer edges). Finally, if the graph has treewidth at least $k$, then the last $k+1$ vertices can be eliminated in any order so we can terminate execution of the algorithm earlier.

We note that our algorithm does not actually compute a tree decomposition or elimination order, but could easily be modified to do so. Currently, the algorithm stores with each (partial) solution one additional integer, which indicates which four vertices were the last to be eliminated. To reconstruct the solution, one could either store a copy of (one in every four of) the output lists on the disk, or repeatedly add the last four vertices to $C$ and rerun the algorithm to obtain the next four vertices (with each iteration taking less time than the previous, since the size of $C$ has increased).

\subsection{Duplicate Elimination using Bloom Filters}

Each set $S$ may be generated in multiple ways by adding different vertices to subsets $S'\subseteq S$; if we do not detect whether a set $S$ is already in the output list when adding it, we risk the algorithm generating $\Omega(n!)$ sets. To detect whether a set $S$ is already in the output, we use a Bloom filter \cite{bloom}: Bloom filters are a classical data structure in which an array $A$ of $m$ bits can be used to encode the presence of $n$ elements by means of $k$ hash functions. To insert an element $S$, we compute $k$ independent hash functions $\{H_i | 1\leq i \leq k\}$ each of which indicates one position in the array, $A[H_i(S)]$, which should be set to $1$. If any of these bits was previously zero, then the element was not yet present in the filter, and otherwise, the probability of a false positive is approximately $(1-e^{-kn/m})^k$.

In our implementation, we compute two 32-bit hashes $h_1(S),h_2(S)$ using Murmur3 \cite{murmur3}, which we then combine linearly to obtain hashes $H_i(S)=h_1(S) + i\cdot h_2(S)$ (which is nearly as good as using $k$ independent hash functions \cite{bloomfilter}).

In our experiments, we have used $\frac{m}{n} \geq 24$ and $k=17$ to obtain a low (theoretical) false positive probability of around $1$ in $100.000$. We note that the possibility of false positives results in a Monte Carlo algorithm (the algorithm may inadvertently decide that the treewidth is higher than it really is). Indeed, given that many millions of states are generated during the search we are guaranteed that the Bloom filter will return some false positives, however, this does not immediately lead to incorrect results: it is still quite unlikely that all of the states leading to an optimal solution are pruned, since there are often multiple feasible elimination orders.

The Bloom filter is very suitable for implementation on a GPU, since our target architecture (and indeed, most GPUs) offers a very fast atomic OR operation \cite{fermi}. We note that addressing a Bloom filter concurrently may also introduce false negatives if multiple threads attempt to insert the same element simultaneously. To avoid this, we use the initial hash value to pick one of 65.536 mutexes to synchronize access (this allows most operations to happen wait-free, and only a collision on the initial hash value causes one thread to wait for another).

\subsection{Minor-Min-Width}

Search algorithms for treewidth are often enhanced with various heuristics and pruning rules to speed up the computation. One very popular choice (used by e.g. \cite{gogate,FPBB,bestTW}) is minor-min-width (MMW) \cite{gogate} (also known as MMD+(min-d)) \cite{lowerbounds}). MMW is based on the observation that the minimum degree of a vertex is a lower bound on the treewidth, and that contracting edges (i.e. taking minors) does not increase the treewidth. MMW repeatedly selects a minimum degree vertex, and then contracts it with a neighbor of minimum degree, in an attempt to obtain a minor with large minimum degree (if we encounter a minimum degree that exceeds our target treewidth, we know that we can discard the current state). As a slight improvement to this heuristic, the second smallest vertex degree is also a lower bound on the treewidth \cite{lowerbounds}.

Given a subset $S\subseteq G$, we would like to compute the treewidth of the graphs that remains after eliminating $S$ from $G$. The most straightforward method is to explicitly create a copy of $G$, eliminate the vertices of $S$, and then repeatedly perform the contraction as described above. However, storing e.g. an adjacency list representation of these intermediate graphs would exceed the available shared memory and size of the caches. As we would like to avoid transferring large amounts of data to and from global memory, we implemented a method to compute MMW without explicitly storing the intermediate graphs.

Our algorithm tracks the current degrees of the vertices (which, conveniently, we already have computed to determine which vertices can be eliminated). It is thus easy to select a minimum degree vertex $v$. Since we do not know what vertices it is adjacent to (in the intermediate graph), we must select a minimum degree neighbor by using a depth-first search, similarly to how we compute the vertex degrees in Listing \ref{lst:basicalgo}. Once we have found a minimum degree neighbor $u$, we run a second dept-first search to compute the number of neighbors $u$ has in common with $v$, allowing us to update the degree of $v$. To keep track of which vertices have been contracted, we use a disjoint set data structure.

The disjoint set structure and list of vertex degrees together use only two bytes per vertex (for a graph of up to 256 vertices), thus, they fit our memory constraints whereas an adjacency matrix or adjacency list (for dense graphs, noting that the graphs in question can quickly become dense as vertices are eliminated) would readily exceed it.

\section{Experiments}

\subsection{Instances}

We selected a number of instances from the PACE 2016 dataset \cite{PACEcall} and libtw \cite{libtw}.

All instances were preprocessed using the preprocessing rules of our PACE submission \cite{BZTreewidth}, which split the graph using \emph{safe} separators: we first split the graph into its connected components, then split on articulation points, then on articulation pairs (making the remaining components 3-connected) and finally - if we can establish that this is safe - on articulation triplets (resulting in the 4-connected components of the graph). We then furthermore try to detect (almost) clique separators in the graph, and split on those. For a more detailed treatment of these preprocessing rules, we refer to \cite{safeseps}.

\subsection{General Benchmark}

We first present an experimental evaluation of our algorithm (without using MMW) on a set of benchmark graphs. Table \ref{tab:exp1} shows the number of vertices, computed treewidth, time taken (in seconds) on the GPU and the number of sets $S$ explored. Note that the time does not include the time taken for preprocessing, and that the vertex count is that of the preprocessed graph (and thus, the original graph may have been larger).

\begin{table}[h]
\centering \texttt{
\begin{tabular}{ | l | c | c | r | r | r | }
  \hline			
Name & $|V|$ & tw & \multicolumn{2}{c|}{Time (sec.)} & Exp  \\
  & & & GPU & CPU & \\
  \hline
1e0b\_graph	&	55	&	24	&	\gtime{778828}	&	-	&	\nexp{1731299698}	\\	\hline
1fjl\_graph*	&	57	&	26	&	\gtime{1734559}	&	-	&	\nexp{3682951713}	\\	\hline	
1igd\_graph	&	59	&	25	&	\gtime{106808}	&	\ctime{5116568}	&	\nexp{260659220}	\\	\hline
1ku3\_graph	&	60	&	22	&	\gtime{234853}	&	-	&	\nexp{542195037}	\\	\hline
1ubq*	&	47	&	\st{11}	&	\gtime{1132360}	&	-	&	\nexp{2296029372}	\\	\hline
8x6\_torusGrid*	&	48	&	\st{7}	&	\gtime{1106367}	&	-	&	\nexp{2098741672}	\\	\hline
BN\_97	&	48	&	18	&	\gtime{1024056}	&	-	&	\nexp{2306775237}	\\	\hline
BN\_98	&	47	&	21	&	\gtime{689341}	&	-	&	\nexp{1589013379}	\\	\hline
contiki\_dhcpc\_handle\_dhcp*	&	39	&	6	&	\gtime{1490413}	&	-	&	\nexp{2933148926}	\\	\hline
DoubleStarSnark	&	30	&	6	&	\gtime{34497}	&	\ctime{873370}	&	\nexp{87628083}	\\	\hline
DyckGraph	&	32	&	7	&	\gtime{280488}	&	-	&	\nexp{638882602}	\\	\hline
HarborthGraph*	&	40	&	5	&	\gtime{697909}	&	-	&	\nexp{1535664481}	\\	\hline
KneserGraph\_8\_3*	&	56	&	\st{24}	&	\gtime{1711441}	&	-	&	\nexp{4125978014}	\\	\hline
McGeeGraph	&	24	&	7	&	\gtime{1298}	&	\ctime{25267}	&	\nexp{3853249}	\\	\hline
myciel4	&	23	&	10	&	\gtime{234}	&	\ctime{460}	&	\nexp{97765}	\\	\hline
myciel5*	&	47	&	19	&	\gtime{2000829}	&	\ctime{70608054}	&	\nexp{4003611855}	\\	\hline
NonisotropicUnitaryPolarGraph\_3\_3	&	63	&	53	&	\gtime{1158}	&	\ctime{60444}	&	\nexp{1558852}	\\	\hline
queen5\_5	&	25	&	18	&	\gtime{212}	&	\ctime{23}	&	\nexp{3134}	\\	\hline
queen6\_6	&	36	&	25	&	\gtime{254}	&	\ctime{389}	&	\nexp{35973}	\\	\hline
queen7\_7	&	49	&	35	&	\gtime{966}	&	\ctime{43491}	&	\nexp{1901029}	\\	\hline
queen8\_8	&	64	&	45	&	\gtime{26284}	&	\ctime{2044470}	&	\nexp{57892754}	\\	\hline
RandomBarabasiAlbert\_100\_2*	&	41	&	12	&	\gtime{1609641}	&	-	&	\nexp{3283134936}	\\	\hline
RandomBoundedToleranceGraph\_60	&	59	&	30	&	\gtime{274}	&	\ctime{635}	&	\nexp{56028}	\\	\hline
SylvesterGraph	&	36	&	15	&	\gtime{247921}	&	-	&	\nexp{631663187}	\\	\hline
te*	&	62	&	\st{7}	&	\gtime{1171702}	&	-	&	\nexp{2163144692}	\\	\hline
water	&	21	&	9	&	\gtime{197}	&	\ctime{6}	&	\nexp{1240}	\\	\hline
\end{tabular}}
\caption{Performance of the algorithm on several benchmark graphs, using global memory and a work size of 128.}
\label{tab:exp1}
\end{table}

The size of the input and output lists were limited by the memory available on our GPU. With the current configuration (limited to graphs of at most 64 vertices - though the code is written to be flexible and can easily be changed to support up to $256$ vertices), these lists could hold at most 180 million states (i.e., subsets $S\subseteq V$ that have a feasible partial elimination order) each. If at any iteration this number was exceeded, the excess states were discarded. The algorithm was allowed to continue execution for the current treewidth $k$, but was terminated when trying the next higher treewidth (since we might have discarded a state that would have lead to a solution with treewidth $k$, the answer would no longer be exact). The states were the capacity of the lists was exceed are marked with \texttt{*}, if the algorithm was terminated then the treewidth is stricken through (and represents the candidate value for treewidth at which the algorithm was terminated, and \emph{not} the treewidth of the graph, which is likely higher).

For instance, for graph \texttt{1ubq} the capacity of the lists was first exceeded at treewidth $10$, and the algorithm was terminated at treewidth $11$ (and thus the actual treewidth is at least $10$, but likely higher). For graph \texttt{myciel5}, the capacity of the lists was first exceeded at treewidth $19$, but still (despite discarding some states) a solution of treewidth $19$ was nevertheless found (which we thus know is the exact treewidth).

For several graphs (those where the GPU version of the algorithm took at most 5 minutes), we also benchmarked a sequential version of the same algorithm on the CPU. In some cases, the algorithm achieves a very large speedup compared to the CPU version (up to $77\times$, in the case of \texttt{queen8\_8}). Additionally, for \texttt{myciel5}, we also ran the CPU-based algorithm, which took more than 19 hours to finish. The GPU version only took 34 minutes.

The GPU algorithm can process a large amount of states in a very short time. For example, for the graph \texttt{1fjl}, 3680 million states were explored in just 1730 seconds, i.e., over 2 million states were processed each second (and for each state, a $\Theta(|V||E|)$-time algorithm is executed). The highest throughput (2.5 million states/sec.) is achieved on \texttt{SylvesterGraph}, but this graph has relatively few vertices.

We caution the reader that the graph names are somewhat ambiguous. For instance, the \texttt{queen7\_7} instance is from libtw and has treewidth 35. The 2016 PACE instances include a graph called \texttt{dimacs\_queen7\_7} which only has treewidth 28. The instances used in our evaluation are available from our GitHub repository \cite{github}.

\subsection{Work Size and Global v.s. Shared Memory}\label{sec:memory}

In this section, we study the effect of work size and whether shared or global memory is used on the running time of our implementation.

Recall that shared memory is a small amount (in our case, 96KiB) of memory that is physically close to each Streaming Multiprocessor, and is therefore in principle faster than the (much larger, off-chip) global memory. We would therefore expect that our implementation is faster when used with shared memory.

Each SM contains $128$ CUDA cores, and thus $4$ warps of $32$ threads each can be executed simultaneously on each SM. The work size (which should be a multiple of $32$), represents the number of threads we assign to each SM. If we set the work size larger than $128$, more threads than can physically be executed at once are assigned to one SM. The SM can then switch between executing different warps, for instance to hide latency of memory accesses. If the work size is smaller than $128$, a number of CUDA cores will be unutilized.

In Table \ref{tab:memwork}, we present some experiments that show running times on several graphs, depending on whether shared memory or global memory is used, for several sizes of work group (which is the number of threads allocated to a single SM).

There is not much difference between running the program using shared or global memory. In most instances, the shared memory version is slightly faster. Surprisingly, it also appears that the work size used does not affect the running time significantly. This suggests that our program is limited by the throughput of memory, rather than being computationally-bound.

\begin{table}[h]
\centering \texttt{
\makebox[\textwidth][c]{\begin{tabular}{ | l | c | c | r | r | r | r | }
  \hline			
Name & $|V|$ & tw & \multicolumn{4}{c|}{Time (sec.)}  \\
  & & & $W=32$ & $W=64$ & $W=128$ & $W=256$  \\ \hline
1igd\_graph (G)	&	59	&	25	&	\gtime{109180}	&	\gtime{106851}	&	\gtime{106808}	&	\gtime{106551}	\\	\hline
1igd\_graph (S)	&	59	&	25	&	\gtime{94760}	&	\gtime{95572}	&	\gtime{98187}	&	\gtime{102615}	\\	\hline
1ku3\_graph (G)	&	60	&	22	&	\gtime{238479}	&	\gtime{235108}	&	\gtime{234854}	&	\gtime{234535}	\\	\hline
1ku3\_graph (S)	&	60	&	22	&	\gtime{213549}	&	\gtime{216710}	&	\gtime{222473}	&	\gtime{229839}	\\	\hline
DoubleStarSnark (G)	&	30	&	6	&	\gtime{34259}	&	\gtime{34510}	&	\gtime{34498}	&	\gtime{34504}	\\	\hline
DoubleStarSnark (S)	&	30	&	6	&	\gtime{32819}	&	\gtime{32810}	&	\gtime{32832}	&	\gtime{32871}	\\	\hline
DyckGraph (G)	&	32	&	7	&	\gtime{277888}	&	\gtime{280513}	&	\gtime{280488}	&	\gtime{280500}	\\	\hline
DyckGraph (S)	&	32	&	7	&	\gtime{266051}	&	\gtime{265918}	&	\gtime{266271}	&	\gtime{266601}	\\	\hline
NonisotropicUnitaryPolarGraph\_3\_3 (G)	&	63	&	53	&	\gtime{1262}	&	\gtime{1148}	&	\gtime{1158}	&	\gtime{1172}	\\	\hline
NonisotropicUnitaryPolarGraph\_3\_3 (S)	&	63	&	53	&	\gtime{1053}	&	\gtime{1017}	&	\gtime{1030}	&	\gtime{1065}	\\	\hline
queen7\_7 (G)	&	49	&	35	&	\gtime{1036}	&	\gtime{969}	&	\gtime{967}	&	\gtime{977}	\\	\hline
queen7\_7 (S)	&	49	&	35	&	\gtime{860}	&	\gtime{861}	&	\gtime{876}	&	\gtime{890}	\\	\hline
queen8\_8 (G)	&	64	&	45	&	\gtime{29527}	&	\gtime{26567}	&	\gtime{26285}	&	\gtime{25991}	\\	\hline
queen8\_8 (S)	&	64	&	45	&	\gtime{25052}	&	\gtime{24057}	&	\gtime{24523}	&	\gtime{25030}	\\	\hline
\end{tabular}}}
\caption{Running time (sec.) for various work group sizes ($W$), using shared (S) or global (G) memory. Each cell lists the average result of 4 test runs, where the complete set of runs was executed in a randomized order.}
\label{tab:memwork}
\end{table}

\begin{table}[h]
\centering \texttt{
\begin{tabular}{ | l | c | c | r | r | r | }
  \hline			
Name & $|V|$ & tw & \multicolumn{2}{c|}{Time (sec.)} & Exp  \\
  & & & GPU & CPU & \\
  \hline
1e0b\_graph	&	55	&	24	&	\gtime{720504}	&	-	&	\nexp{1731299697}	\\	\hline
1fjl\_graph*	&	57	&	26	&	\gtime{1595005}	&	-	&	\nexp{3656827840}	\\	\hline	
1igd\_graph	&	59	&	25	&	\gtime{98186}	&	\ctime{5116568}	&	\nexp{260659220}	\\	\hline
1ku3\_graph	&	60	&	22	&	\gtime{222473}	&	-	&	\nexp{542195037}	\\	\hline
1ubq*	&	47	&	\st{11}	&	\gtime{1038873}	&	-	&	\nexp{2290803382}	\\	\hline
8x6\_torusGrid*	&	48	&	\st{7}	&	\gtime{1039769}	&	-	&	\nexp{2077190472}	\\	\hline
BN\_97	&	48	&	18	&	\gtime{944285}	&	-	&	\nexp{2306775238}	\\	\hline
BN\_98	&	47	&	21	&	\gtime{642902}	&	-	&	\nexp{1589013379}	\\	\hline
contiki\_dhcpc\_handle\_dhcp*	&	39	&	6	&	\gtime{1353034}	&	-	&	\nexp{2835417012}	\\	\hline
DoubleStarSnark	&	30	&	6	&	\gtime{32832}	&	\ctime{873370}	&	\nexp{87628083}	\\	\hline
DyckGraph	&	32	&	7	&	\gtime{266271}	&	-	&	\nexp{638882602}	\\	\hline
HarborthGraph*	&	40	&	5	&	\gtime{646916}	&	-	&	\nexp{1532611838}	\\	\hline
KneserGraph\_8\_3*	&	56	&	\st{24}	&	\gtime{1580621}	&	-	&	\nexp{4103569889}	\\	\hline
McGeeGraph	&	24	&	7	&	\gtime{1235}	&	\ctime{25267}	&	\nexp{3853249}	\\	\hline
myciel4	&	23	&	10	&	\gtime{238}	&	\ctime{460}	&	\nexp{97765}	\\	\hline
myciel5*	&	47	&	19	&	\gtime{1845884}	&	\ctime{70608054}	&	\nexp{3994710744}	\\	\hline
NonisotropicUnitaryPolarGraph\_3\_3	&	63	&	53	&	\gtime{1029}	&	\ctime{60444}	&	\nexp{1558852}	\\	\hline
queen5\_5	&	25	&	18	&	\gtime{179}	&	\ctime{23}	&	\nexp{3134}	\\	\hline
queen6\_6	&	36	&	25	&	\gtime{241}	&	\ctime{389}	&	\nexp{35973}	\\	\hline
queen7\_7	&	49	&	35	&	\gtime{875}	&	\ctime{43491}	&	\nexp{1901029}	\\	\hline
queen8\_8	&	64	&	45	&	\gtime{24522}	&	\ctime{2044470}	&	\nexp{57892754}	\\	\hline
RandomBarabasiAlbert\_100\_2*	&	41	&	12	&	\gtime{1473373}	&	-	&	\nexp{3264359373}	\\	\hline
RandomBoundedToleranceGraph\_60	&	59	&	30	&	\gtime{263}	&	\ctime{635}	&	\nexp{56028}	\\	\hline
SylvesterGraph	&	36	&	15	&	\gtime{229185}	&	-	&	\nexp{631663187}	\\	\hline
te*	&	62	&	\st{7}	&	\gtime{1098076}	&	-	&	\nexp{2138569876}	\\	\hline
water	&	21	&	9	&	\gtime{207}	&	\ctime{6}	&	\nexp{1240}	\\	\hline
\end{tabular}}
\caption{The same experiment as in Table \ref{tab:exp1}, but using shared instead of global memory. Work size 128.}
\end{table}

\subsection{Minor-Min-Width}

In Table \ref{tab:mmw}, we list results obtained when using Minor-Min-Width to prune states.

\begin{table}[h]
\centering \texttt{
\makebox[\textwidth][c]{\begin{tabular}{ | l | c | c | r | r | r | r | r | }
  \hline			
Name & $|V|$ & tw & \multicolumn{2}{c|}{With MMW} & \multicolumn{2}{c|}{Without MMW}  \\
  & & & Time & Exp & Time & Exp \\
  \hline
1e0b\_graph	&	55	&	24	&	\gtime{2751198}	& \nexp{1657391747}	&	\gtime{778828}	&	\nexp{1731299698}	\\	\hline
1fjl\_graph*	&	57	&	26	&	\texttt{timeout}	& \nexp{3259473749}	&	\gtime{1734559}	&	\nexp{3682951713}	\\	\hline	
1igd\_graph	&	59	&	25	&	\gtime{471319}	& \nexp{235151811}	&	\gtime{106808}	&	\nexp{260659220}	\\	\hline
1ku3\_graph	&	60	&	22	&	\gtime{1089498}	& \nexp{510573831}	&	\gtime{234853}	&	\nexp{542195037}	\\	\hline
1ubq*	&	47	&	\st{11}	&	\gtime{2011078}	& \nexp{1503291328}	&	\gtime{1132360}	&	\nexp{2296029372}	\\	\hline
8x6\_torusGrid*	&	48	&	\st{7}	&	\gtime{1352593}	& \nexp{1299664022}	&	\gtime{1106367}	&	\nexp{2098741672}	\\	\hline
BN\_97	&	48	&	18	&	\gtime{2260574}	& \nexp{2019966580}	&	\gtime{1024056}	&	\nexp{2306775237}	\\	\hline
BN\_98	&	47	&	21	&	\gtime{1481937}	& \nexp{1436839298}	&	\gtime{689330}	&	\nexp{1589013379}	\\	\hline
contiki\_dhcpc\_handle\_dhcp*	&	39	&	6	&	\gtime{2672220}	& \nexp{2896176997}	&	\gtime{1490413}	&	\nexp{2933148926}	\\	\hline
DoubleStarSnark	&	30	&	6	&	\gtime{38295}	& \nexp{76042513}	&	\gtime{34497}	&	\nexp{87628083}	\\	\hline
DyckGraph	&	32	&	7	&	\gtime{343311}	& \nexp{592154794}	&	\gtime{280488}	&	\nexp{638882602}	\\	\hline
HarborthGraph*	&	40	&	5	&	\gtime{1456102}	& \nexp{1571045413}	&	\gtime{1711441}	&	\nexp{1535664481}	\\	\hline
KneserGraph\_8\_3*	&	56	&	\st{24}	&	\gtime{1334084}	& \nexp{1221365732}	&	\gtime{1732752}	&	\nexp{4128586629}	\\	\hline
McGeeGraph	&	24	&	7	&	\gtime{1875}	& \nexp{3417125}	&	\gtime{1298}	&	\nexp{3853249}	\\	\hline
myciel4	&	23	&	10	&	\gtime{614}	& \nexp{75141}	&	\gtime{234}	&	\nexp{97765}	\\	\hline
myciel5*	&	47	&	19	&	\gtime{2546413}	& \nexp{3196230168}	&	\gtime{2000829}	&	\nexp{4003611855}	\\	\hline
NonisotropicUnitaryPolarGraph\_3\_3	&	63	&	53	&	\gtime{3359}	& \nexp{1298020}	&	\gtime{1158}	&	\nexp{1558852}	\\	\hline
queen5\_5	&	25	&	18	&	\gtime{810}	& \nexp{2909}	&	\gtime{212}	&	\nexp{3134}	\\	\hline
queen6\_6	&	36	&	25	&	\gtime{1155}	& \nexp{30833}	&	\gtime{254}	&	\nexp{35973}	\\	\hline
queen7\_7	&	49	&	35	&	\gtime{2909}	& \nexp{1749860}	&	\gtime{966}	&	\nexp{1901029}	\\	\hline
queen8\_8	&	64	&	45	&	\gtime{83471}	& \nexp{51094313}	&	\gtime{26284}	&	\nexp{57892754}	\\	\hline
RandomBarabasiAlbert\_100\_2*	&	41	&	12	&	\gtime{2387180}	& \nexp{2843806917}	&	\gtime{1609641}	&	\nexp{3283134936}	\\	\hline
RandomBoundedToleranceGraph\_60	&	59	&	30	&	\gtime{630}	& \nexp{47754}	&	\gtime{274}	&	\nexp{56028}	\\	\hline
SylvesterGraph	&	36	&	15	&	\gtime{274186}	& \nexp{502854646}	&	\gtime{247921}	&	\nexp{631663187}	\\	\hline
te*	&	62	&	\st{10}	&	\gtime{2256440}	& \nexp{1694586333}	&	\gtime{1171702}	&	\nexp{2163144692}	\\	\hline
water	&	21	&	9	&	\gtime{410}	& \nexp{938}	&	\gtime{197}	&	\nexp{1240}	\\	\hline
\end{tabular}}}
\caption{The effect of using the Minor-Min-Width Heuristic. Time is in seconds. Global memory, work size 128.}
\label{tab:mmw}
\end{table}

\begin{table}[h]
\centering \texttt{
\makebox[\textwidth][c]{\begin{tabular}{ | l | c | c | r | r | r | r | r | }
  \hline			
Name & $|V|$ & tw & \multicolumn{2}{c|}{With MMW} & \multicolumn{2}{c|}{Without MMW}  \\
  & & & Time & Exp & Time & Exp \\
  \hline
1e0b\_graph	&	55	&	24	&	\gtime{3648754}	& \nexp{1657391746}	&	\gtime{720504}	&	\nexp{1731299697}	\\	\hline
1fjl\_graph*	&	57	&	26	&	\texttt{timeout}	& \nexp{2440028938}	&	\gtime{1595005}	&	\nexp{3656827840}	\\	\hline	
1igd\_graph	&	59	&	25	&	\gtime{645167}	& \nexp{235151811}	&	\gtime{98186}	&	\nexp{260659220}	\\	\hline
1ku3\_graph	&	60	&	22	&	\gtime{1462862}	& \nexp{510573831}	&	\gtime{222473}	&	\nexp{542195037}	\\	\hline
1ubq*	&	47	&	\st{11}	&	\gtime{2511222}	& \nexp{1484913654}	&	\gtime{1038873}	&	\nexp{2290803382}	\\	\hline
8x6\_torusGrid*	&	48	&	\st{7}	&	\gtime{1889420}	& \nexp{1306203955}	&	\gtime{1039769}	&	\nexp{2077190472}	\\	\hline
BN\_97	&	48	&	18	&	\gtime{3127862}	& \nexp{2019966557}	&	\gtime{944285}	&	\nexp{2306775238}	\\	\hline
BN\_98	&	47	&	21	&	\gtime{1973437}	& \nexp{1436839298}	&	\gtime{642902}	&	\nexp{1589013379}	\\	\hline
contiki\_dhcpc\_handle\_dhcp*	&	39	&	6	&	\gtime{3266014}	& \nexp{2859142248}	&	\gtime{1353034}	&	\nexp{2835417012}	\\	\hline
DoubleStarSnark	&	30	&	6	&	\gtime{50853}	& \nexp{76042513}	&	\gtime{32832}	&	\nexp{87628083}	\\	\hline
DyckGraph	&	32	&	7	&	\gtime{439804}	& \nexp{592154794}	&	\gtime{266271}	&	\nexp{638882602}	\\	\hline
HarborthGraph*	&	40	&	5	&	\gtime{1895588}	& \nexp{1559260727}	&	\gtime{646916}	&	\nexp{1532611838}	\\	\hline
KneserGraph\_8\_3*	&	56	&	\st{24}	&	\gtime{1875589}	& \nexp{1210987306}	&	\gtime{1580621}	&	\nexp{4103569889}	\\	\hline
McGeeGraph	&	24	&	7	&	\gtime{2015}	& \nexp{3417125}	&	\gtime{1235}	&	\nexp{3853249}	\\	\hline
myciel4	&	23	&	10	&	\gtime{955}	& \nexp{75141}	&	\gtime{238}	&	\nexp{97765}	\\	\hline
myciel5*	&	47	&	19	&	\gtime{3366323}	& \nexp{3182573127}	&	\gtime{1845884}	&	\nexp{3994710744}	\\	\hline
NonisotropicUnitaryPolarGraph\_3\_3	&	63	&	53	&	\gtime{4173}	& \nexp{1298020}	&	\gtime{1029}	&	\nexp{1558852}	\\	\hline
queen5\_5	&	25	&	18	&	\gtime{704}	& \nexp{2909}	&	\gtime{179}	&	\nexp{3134}	\\	\hline
queen6\_6	&	36	&	25	&	\gtime{998}	& \nexp{30833}	&	\gtime{241}	&	\nexp{35973}	\\	\hline
queen7\_7	&	49	&	35	&	\gtime{3636}	& \nexp{1749860}	&	\gtime{837}	&	\nexp{1901029}	\\	\hline
queen8\_8	&	64	&	45	&	\gtime{115984}	& \nexp{51094313}	&	\gtime{24522}	&	\nexp{57892754}	\\	\hline
RandomBarabasiAlbert\_100\_2*	&	41	&	12	&	\gtime{3079744}	& \nexp{2834378395}	&	\gtime{1473373}	&	\nexp{3264359373}	\\	\hline
RandomBoundedToleranceGraph\_60	&	59	&	30	&	\gtime{666}	& \nexp{47754}	&	\gtime{263}	&	\nexp{56028}	\\	\hline
SylvesterGraph	&	36	&	15	&	\gtime{368409}	& \nexp{502854646}	&	\gtime{229185}	&	\nexp{631663187}	\\	\hline
te*	&	62	&	\st{10}	&	\gtime{3097746}	& \nexp{1632566154}	&	\gtime{1098076}	&	\nexp{2138569876}	\\	\hline
water	&	21	&	9	&	\gtime{543}	& \nexp{938}	&	\gtime{207}	&	\nexp{1240}	\\	\hline
\end{tabular}}}
\caption{The same experiment as in Table \ref{tab:mmw}, but using shared instead of global memory. Work size 128.}
\end{table}

The computational expense of using MMW is comparable to that of the initial computation (for determining the degree of vertices): the algorithm does a linear search for a minimum degree vertex (using the precomputed degree values), and then does a graph traversal (using BFS) to find a minimum degree neighbour (recall that we do not store the intermediate graph, and use only a single copy of the original graph). Once such a neighbour is found, the contraction is performed (by updating the disjoint set data structure) and another graph traversal is required (to compute the number of common neighbours, and thus update the degree of the vertex).

The lower bound given by MMW does not appear to be very strong, at least for the graphs considered in our experiment: the reduction in number of states expanded is not very large (for instance, from 1730 million states to 1660 million for \texttt{1e0b}, or from 1590 million to 1480 million for \texttt{BN\_98}). The largest reductions are visible for graphs on which we run out of memory (for instance, from 4130 million to 1330 million for \texttt{KneserGraph\_8\_3}), but this is likely because the search is terminated before we reach the actual treewidth (so we avoid the part of our search where using a heuristic is least effective) and there are no graphs on which we previously ran out of memory for which MMW allows us to determine the treewidth (the biggest improvement is that we are able to determine that \texttt{te} has treewidth at least 10, up from treewidth at least 7).

Consistent with the relatively low reduction in the number of states expanded, we see the computation using MMW typically takes around $2-3$ times longer. On the graphs considered here, the reduction in search space offered by MMW does not offset the additional cost of computing it.

Again, the GPU version is significantly faster than executing the same algorithm on the CPU: we observed a $55\times$ speedup for \texttt{queen8\_8}. Still, given what we observed in Section \ref{sec:memory}, it is not clear whether our approach of not storing the intermediate graphs explicitly is indeed the best approach. Our main motivation for taking this approach was to be able to store the required data structures entirely in shared memory, but our experiments indicate that for MMW, using global memory gives better performance than using shared memory. However, the relatively good performance of global memory might be (partially) due to caching and the small amount of data transferred, so it is an interesting open question to determine whether the additional memory costs of using more involved data structures is compensated by the potential speedup.

\subsection{Loop Unnesting}

Finally, we experimented with another technique, which aims to increase parallelism (and thus speedup) by limiting branch divergence. However, as the results were discouraging, we limit ourselves to a brief discussion.

The algorithm of Listing \ref{lst:basicalgo} consists of a loop (lines 5--22) over the (not yet eliminated) vertices, inside of which is a depth-first search (which computes the degree of the vertex, to determine whether it can be eliminated). The depth-first search in turn consists of a loop which runs until the stack becomes empty (lines 10--19) inside of which is a final loop over the neighbours of the current vertex (lines 12--18). This leads to two sources of branch divergence:

\begin{itemize}
\item First, if the graph is irregular, all threads in a warp have to wait for the thread that is processing the highest degree vertex, even if they only have low-degree vertices.

\item Second, all threads in a warp have to wait for the longest of the BFS searches to finish before they can start processing the next vertex.
\end{itemize}

To alleviate this, we proposed a technique which we call \emph{loop unnesting}: rather than have 3 nested loops, we have only one loop, which simulates a state machine with 3 states: (1) processing the adjacency list of a vertex, (2) having finished processing of an adjacency list and being ready to pop a new vertex off the queue, or (3) having finished a BFS, and being ready to begin computing the degree of a new vertex.

We considered a slightly more general version of this idea: in an $(x,y)$-unnesting of our program, after every $x$ iterations of the inner loop (exploring neighbours of the current vertex) one iteration of the middle loop is executed (if exploring the adjacency list is finished, get a new vertex from the queue), and for every $y$ iterations of the middle loop, one iteration of the outer loop is executed (begin processing an entirely new vertex). Thus, a $(1,1)$-unrolling corresponds to the state machine simulation described above, and an $(\infty,\infty)$-unrolling corresponds to the original program.

Picking the right values for $x,y$ means finding the right trade-off between checking frequently enough whether a thread is ready to start working on another vertex, and the cost of performing those checks. What we observed was surprising: while $(1,1)$, $(3,2)$ and $(1,\infty$)-unrollings gave reasonable results, the best results were obtained with $(\infty,\infty)$-unrollings (i.e. the original, unmodified algorithm) and the performance of $(\infty,1)$-unrollings was abysmal.

We believe that a possible explanation may be that loop unnesting does work to some extent, but not unnesting the loops has the advantage that all BFS searches running simultaneously start from the same initial vertex, and (up to differences caused by different sets $S$ being used) will access largely the same values from the adjacency lists at the same time, which may increase the efficiency of read operations. On the other hand, $(\infty,1)$-unnesting can not take advantage of either phenomenon: different initial vertices may be processed at any given time (so there is little consistency in memory accesses) and the inner loop is not unnested at all so there is no potential to gain speedup there either. Perhaps for larger graphs, where the difference in length of adjacency lists may be more pronounced, or the amount of time a BFS takes varies more strongly with the initial vertes and $S$, loop unnesting does provide speed up, but for the graphs considered here it does not appear to be a beneficial choice.

\begin{table}[h]
\centering \texttt{
\makebox[\textwidth][c]{\begin{tabular}{ | l | c | c | r | r | r | r | r | }
  \hline			
Name & $|V|$ & tw & \multicolumn{5}{c|}{Time (sec.)}  \\
  & & & \centering $(1,1)$ & \centering $(\infty,\infty)$ & \centering $(1,\infty)$ & \centering $(\infty,1)$ &  $(3,2)$ \\
  \hline
1igd\_graph (G)	&	59	&	25	&	\gtime{121357}	&	\gtime{106808}	&	\gtime{113004}	&	\gtime{313235}	&	\gtime{111967}		\\	\hline
1igd\_graph (S)	&	59	&	25	&	\gtime{128274}	&	\gtime{98187}	&	\gtime{105476}	&	\gtime{241259}	&	\gtime{114323}		\\	\hline
1ku3\_graph (G)	&	60	&	22	&	\gtime{267997}	&	\gtime{234854}	&	\gtime{246517}	&	\gtime{626457}	&	\gtime{247705}		\\	\hline
1ku3\_graph (S)	&	60	&	22	&	\gtime{291454}	&	\gtime{222473}	&	\gtime{237765}	&	\gtime{490067}	&	\gtime{262329}		\\	\hline
DoubleStarSnark (G)	&	30	&	6	&	\gtime{35339}	&	\gtime{34498}	&	\gtime{34573}	&	\gtime{38757}	&	\gtime{34676}		\\	\hline
DoubleStarSnark (S)	&	30	&	6	&	\gtime{32881}	&	\gtime{32832}	&	\gtime{32867}	&	\gtime{37672}	&	\gtime{32790}		\\	\hline
DyckGraph (G)	&	32	&	7	&	\gtime{288581}	&	\gtime{280488}	&	\gtime{280907}	&	\gtime{316391}	&	\gtime{282612}		\\	\hline
DyckGraph (S)	&	32	&	7	&	\gtime{266403}	&	\gtime{266271}	&	\gtime{266597}	&	\gtime{305664}	&	\gtime{265775}		\\	\hline
NonisotropicUnitaryPolarGraph\_3\_3 (G)	&	63	&	53	&	\gtime{2171}	&	\gtime{1158}	&	\gtime{1512}	&	\gtime{7107}	&	\gtime{1729}		\\	\hline
NonisotropicUnitaryPolarGraph\_3\_3 (S)	&	63	&	53	&	\gtime{2173}	&	\gtime{1030}	&	\gtime{1399}	&	\gtime{5164}	&	\gtime{1681}		\\	\hline
queen7\_7 (G)	&	49	&	35	&	\gtime{1368}	&	\gtime{967}	&	\gtime{1126}	&	\gtime{4514}	&	\gtime{1197}		\\	\hline
queen7\_7 (S)	&	49	&	35	&	\gtime{1313}	&	\gtime{876}	&	\gtime{1022}	&	\gtime{3301}	&	\gtime{1116}		\\	\hline
queen8\_8 (G)	&	64	&	45	&	\gtime{52219}	&	\gtime{26285}	&	\gtime{34270}	&	\gtime{147538}	&	\gtime{42671}		\\	\hline
queen8\_8 (S)	&	64	&	45	&	\gtime{55441}	&	\gtime{24523}	&	\gtime{33825}	&	\gtime{111470}	&	\gtime{43168}		\\	\hline
\end{tabular}}}
\caption{Results on loop unnesting. Work size used was 128. Each cell lists the average result of 4 test runs, where the complete set of runs was executed in a randomized order.}
\label{tab:unnesting}
\end{table}

\section{Conclusions}

We have presented an algorithm that computes treewidth on the GPU, achieving a very large speedup over running the same algorithm on the CPU. Our algorithm is based on the classical $O^*(2^n)$-time dynamic programming algorithm \cite{ExactAlgos} and our results represent (promising) first steps in speeding up dynamic programming for treewidth on the GPU. The current best known practical algorithm for computing treewidth is the algorithm due to
 Tamaki \cite{tamaki}. This algorithm is much more complicated, and porting it to the GPU would be a formidable challenge but could possibly offer an extremely efficient implementation for computing treewidth.

Given the large speedup achieved, we are no longer mainly limited by computation time. Instead, our ability to solve larger instances is hampered by the memory required to store the very large lists of partial solutions. Using minor-min-width did not prove effective in reducing the number of states considerably, so it would be interesting to see how other heuristics and pruning rules (such as simplicial vertex detection) could be implemented on the GPU.

GPUs are traditionally used to solve easy (e.g. linear time) problems on very large inputs (such as the millions of pixels rendered on a screen, or exploring a graph with millions of nodes), but clearly, the speedup offered by inexpensive GPUs would also be very welcome in solving hard ($\NP$-complete) problems on small instances. Exploring how techniques from FPT and exact algorithms can be used on the GPU raises many interesting problems - not only practical ones, but also theoretical: how should we model complex devices such as GPUs, with their many types of memory and branch divergence issues?

\textbf{Acknowledgements.} We thank Jacco Bikker for discussions on the architecture of GPUs, and Gerard Tel for discussions on hash functions.

\textbf{Source Code and Instances.} We have made our source code, as well as the graphs used for the experiments, available on GitHub \cite{github}.

\bibliography{references}

\end{document}